\documentclass[prl,twocolumn,showpacs,superscriptaddress]{revtex4}
\usepackage{amsmath}
\usepackage{amsfonts}
\usepackage{amssymb}
\usepackage{graphicx}

\begin{document}

\title{Tight multipartite Bell's inequalities involving many measurement settings}
\author{Wies{\l}aw Laskowski}
\affiliation{Instytut Fizyki Teoretycznej i Astrofizyki,
Uniwersytet Gda\'nski, PL-80-952 Gda\'nsk, Poland}
\author{Tomasz Paterek}
\affiliation{Instytut Fizyki Teoretycznej i Astrofizyki,
Uniwersytet Gda\'nski, PL-80-952 Gda\'nsk, Poland}
\author{Marek {\. Z}ukowski}
\affiliation{Instytut Fizyki Teoretycznej i Astrofizyki,
Uniwersytet Gda\'nski, PL-80-952 Gda\'nsk, Poland}
\affiliation{Institut f\"ur Experimentalphysik, Universit\"at
Wien, Boltzmanngasse 5, A--1090 Wien, Austria}
\author{{\v C}aslav Brukner}
\affiliation{Institut f\"ur Experimentalphysik, Universit\"at
Wien, Boltzmanngasse 5, A--1090 Wien, Austria}
\affiliation{Blackett Laboratory, Imperial College London, London SW7 2BW, United Kingdom}
\date{\today}

\begin{abstract}
We derive tight Bell's inequalities for $N\!>\!2$ observers
involving {\em more than two} alternative measurement settings.
We give a {\em necessary} and {\em sufficient} condition for a
general quantum state to violate the new inequalities. The
inequalities are violated by some classes of states, for which {\em
all} standard Bell's inequalities with two measurement settings
per observer are satisfied. 
\end{abstract}


\maketitle

Which quantum states do not allow a local realistic (LR)
description? This question still remains open, mainly because
our present tools to test local realism are not optimal. Most of
Bell's inequalities are for the case in which only {\it two}
measurement settings can be chosen by each observer, e.g. the
Clauser-Horne-Shimony-Holt inequality \cite{CHSH} (CHSH),
inequalities for bipartite higher-dimensional systems \cite{KAZ},
multipartite Bell's inequalities \cite{MERMIN,WZuk,WW,ZB}. One
can call such inequalities ``standard'' ones.

One can expect that allowing the observers to choose between more
than two observables should give {\em more stringent constraints}
on LR models. Thus, new inequalities may {\em extend the class of
non-separable states which cannot be described by LR variables}.
Violation of local realism is an important ingredient for
building quantum information protocols that decrease the
communication complexity \cite{BZJZ} and is a criterion for the
efficient quantum key distribution \cite{SCARANI1}. Multi-setting
Bell's inequalities may lead to such novel schemes.

Although various non-standard Bell's inequalities are
known~\cite{PITOWSKYSVOZIL,ZUK93,GISIN,MASSAR2,COLLINS}, we still
lack an efficient method for their derivation. A
general way is to define the facets of the
correlation polytope \cite{PITOWSKYSVOZIL,COLLINS}. Yet, this is
computationally hard NP-problem \cite{PITOWSKY}. Recently,
Wu and Zong \cite{WZ} derived an inequality for three parties,
which involves {\em four} local settings for two observers and
two settings for the third one, and showed that it is stronger than
the standard three-particle Bell inequalities.

Here, by generalizing the method of Ref.~\cite{WZ}, we derive a
wide class of {\em tight} Bell's inequalities for $N\!>\!2$
parties and many measurement settings. We find a {\em single}
(general) Bell's inequality that generates all inequalities from
this class. We also give a {\em necessary} and {\em sufficient}
condition for the violation of the new inequalities by quantum
predictions. Most importantly, the inequalities reveal violation
of local realism for the classes of states, for which all 
standard Bell's inequalities \cite{WW,ZB} are satisfied.

We start with the case of $N=3$ observers. Suppose that the first
two observers can choose between four settings, and the third one
between two settings. We denote such problem as $4 \times 4
\times 2$. Let $A_i$ with $i \in \{1,2,3,4\}$ stand for the
predetermined local realistic values for the first observer under
the local setting $i$, $B_j$ with $j \in \{1,2,3,4\}$ for similar
values for the second observer, and $C_k$ with $k \in \{1,2\}$
for the values for the third observer (for the given run of the
experiment). We assume that $A_i$, $B_j$ and $C_k$ can take values
$+1$ or $-1$. Then the LR values satisfy the following algebraic
identity:
\begin{equation}
A_{12,S'} \!\equiv \! \hspace{-2mm}\sum_{k,l=1,2}\! \hspace{-2mm}
S'(k,l) (A_1+(-1)^{k}A_2) (B_1+(-1)^lB_2)=\pm 4, \label{kraj}
\end{equation}
where $S'(k,l)$ is {\it any} ``sign" function, i.e. such that
$S'(k,l)\! =\! \pm 1$. Since $|A_i| \!=\! |B_j| \!=\! 1$, only
one term in Eq. (\ref{kraj}) does not vanish, and its value is
$\pm 4$.

In analogous way, one can define $A_{34,S''}$ by replacing
$A_1,A_2,B_1,B_2$ by $A_3,A_4,B_3,B_4$, respectively, and $S'$ by
$S''$. Depending on the value of $m= \pm 1$ one has $(A_{12,S'} +
(-1)^mA_{34,S''})=\pm 8,$ or $0$. By analogy to (\ref{kraj}) one has:
\begin{eqnarray}
 \lefteqn{A_{1234,12} \equiv } \label{GEN} & &  \\ & &  \hspace{-0.4cm} \sum_{k,l=
1,2} \hspace{-2mm}  S(k,l) (A_{12,S'}+(-1)^{k}
A_{34,S''})(C_1+(-1)^{l}C_2)=\pm 16. \nonumber
\end{eqnarray}
After averaging over many runs of the experiment, and introducing
the correlation functions $E_{ijk}\equiv \langle A_i B_j C_k
\rangle_{\textrm{avg}}$ one obtains multisetting Bell's inequalities.
Because of the freedom to choose the sign functions $S, S', S''$,
we have $(2^4)^3=2^{12}$ Bell's inequalities.

We will now show that the family of $2^{12}$ inequalities
can be reduced to a single ``generating'' one. This
inequality will be obtained for non-factorable sign functions $S,
S', S''$. Below we show that a choice of factorable sign function
is equivalent to having a non-factorable one, and some of the
local measurement settings equal. Thus, one does not need to
consider factorable sign functions; one obtains inequalities for 
such sign functions from the generating Bell's
inequality by making some settings equal.

To show this consider function $S(k,l)$ as an example. In general
$S(k,l) \!=\! a(k) + b(k) (-1)^l$, with either $a(k)\!=\!0$ or
$b(k)\! =\! 0$, and $|a(k)| + |b(k)|\! =\ 1$. If $S(k,l)$ is
factorable (i.e. of the form $S(k,l) = s_1(k) s_2(l)$ with
$s_1(k)=s_2(l)=\pm 1$), then either $b(k)\! \equiv \!0$ or
$a(k)\! \equiv \!0$. If we choose, e.g., $b(k)\! \equiv \!0$, then the last factor on the left-hand side of Eq.
(\ref{GEN}) has the following form:
\begin{eqnarray}
\lefteqn{\sum_{l=1,2} S(k,l) (C_1 + (-1)^l C_2) } & & \\
& & = \sum_{l=1,2} (a(k) + b(k) (-1)^{l}) (C_1 + (-1)^{l} C_2)
\\ & & = 2 a(k) C_1 + 2 b(k) C_2 = 2 a(k) C_1. \label{FACTORABLES}
\end{eqnarray}
The setting ``2" for the third observer drops out. 
The final expression (\ref{FACTORABLES}) can be also obtained, for the
non-factorable $S$, by putting $C_1\!=\!C_2$. Further, if one
inserts this result into Eq.~(\ref{GEN}), and, say, $a(k) \!\equiv
\!1$, then after the summation over $k$ the whole term with
settings $3,4$ for the first two observers vanishes. What we get
is a trivial extension of the CHSH inequalities.

The whole family of multisetting Bell's inequalities can be
reduced to one ``generating" inequality which is obtained for $S,
S', S''$ non-factorable. In such cases $a(k) = \pm \frac{1 \pm
(-1)^{k}}{2}$ and $b(k) = \pm \frac{1 \mp (-1)^{k} }{2}$ (the
front signs are free, those in the numerators have to be
different for the two functions). Any other cases are obtainable
by the sign changes $X_i\! \to \!- X_i$ ($X \!=\! A,B,C$). The ``generating" Bell's inequality can be chosen as
\begin{eqnarray}
&& |\langle (C_1 + C_2) [A_1(B_1+B_2) + A_2 (B_1 - B_2)] \label{442_GEN_INEQ} \\
&& + (C_1 - C_2) [A_3(B_3+B_4) + A_4 (B_3 - B_4)]
\rangle_{\textrm{avg}}| \le 4. \nonumber
\end{eqnarray}
Other inequalities can be obtained by making some settings equal.
For example, the $3 \times 3 \times 2$ ones can be
obtained by choosing the settings 1 and 2 identical for the first
two observers.

The method can be generalized for various choices of the number of
parties and the measurement settings. Here we present the
$2^{N-1} \times 2^{N-1} \times 2^{N-2}\times ...
\times 2$ ones. 

Take the case of $N\!=\!4$ observers. We start with the identity (\ref{GEN}). One can introduce a similar
identity for the settings $\{5,6,7,8\}$, for the first two
observers, and $\{3,4\}$, for the third one. The fourth observer
chooses between two settings with LR values $D_1$ and $D_2$.
Applying the same method as before, one obtains an identity which
generates Bell's inequalities of the $8 \times 8 \times 4 \times
2$ type:
\begin{eqnarray}
&&\sum_{k,l = 1,2} \hspace{-2mm} S(k,l) (A_{1234,12}+(-1)^{k}
A_{5678,34}) \nonumber \\
&& \times (D_1 + (-1)^{l} D_2)= \pm 64 \label{8842IDENTITY}
\end{eqnarray}
where $A_{1234,12}$ and  $A_{5678,34}$ depend on some three sign
functions. One may apply this method {\it iteratively},
increasing the number of observers by one, to obtain inequalities
involving exponential (in $N$) number of measurement settings as
given above.

{\em The inequalities are tight.} We give the proof for the case
$4 \times 4 \times 2$ inequalities. It can be adapted to all
inequalities discussed here. The left hand side of the identity
(\ref{GEN}) is equal to $\pm 16$ for any combination of
predetermined LR results. In a 32 dimensional real space, one can
build a convex polytope, containing all possible LR models of the
correlation functions for the specified settings, with vertices
(generators) given by the tensor products of
$v=(A_1,A_2,A_3,A_4)\otimes(B_1,B_2,B_3,B_4)\otimes(C_1,C_2)$. It
has $256$ different vertices. Tight Bell inequalities define the
half-spaces in which is the polytope,  which contain a face of it
in their border hyperplane. If  32  linearly independent vertices
belong to a hyperplane, this hyperplane defines a tight
inequality. Half of the vertices saturate the inequality bounded
by $16$ and  another half saturate  the inequality with $-16$
bound. Every vertex, $v$, saturating the first inequality, has a
partner $-v$, which saturates the other one. Any set of $128$
vertices $v$, which does not contain pairs $v$ and $-v$ contains
a set of 32 linearly independent points (basis). Thus, each
inequality is tight.

{\em The necessary and sufficient condition for violation of
multisetting Bell's inequalities.} To this end we first
rederive the necessary and sufficient condition for violation of
the CHSH inequality by an arbitrary two-qubit
state~\cite{HORODECKI}. The derivation will use certain mathematical ideas that will
be later applied in the analysis of more general cases. We will use a decomposition of general mixed state of
$N$ qubits in terms of the identity operator $\sigma_0=\openone$
in the Hilbert space of individual qubits and the Pauli operators
$\sigma_i$ for three orthogonal directions $i\in \{1,2,3\}$, given 
by $\rho=\frac{1}{2^N} \sum_{k_1,...,k_N=0}^{3}
T_{k_1...k_N} \sigma_{k_1} \otimes ...\otimes \sigma_{k_N}$. The
(real) coefficients $T_{k_1...k_N}$, with $k_j=1,2,3$, form the 
correlation tensor $\hat{T}$.

The full set of inequalities for the $2 \times 2$ problem is
derivable from the identity (\ref{kraj}) and reads:
\begin{equation}
|\langle \sum_{k,l= 1,2} S(k,l) (A_1+(-1)^{k} A_2)(B_1 + (-1)^{l}
B_2) \rangle_{\textrm{avg}}| \le 4. \label{full22}
\end{equation}
The quantum correlation function  $E(\vec A,\vec B)$ is given by
the scalar product of the correlation tensor $\hat T$ with the
tensor product of the local measurement settings represented by
unit vectors $\vec A \otimes \vec B$, i.e. $E(\vec A,\vec B) =
(\vec A \otimes \vec B) \cdot \hat T$. Thus, the condition for a
quantum state endowed with the correlation tensor $\hat T$ to
satisfy the inequality (\ref{full22}), is that for all directions
$ \vec A_1, \vec A_2, \vec B_1, \vec B_2$ one has
\begin{equation}
|\Big[ \sum_{k,l=1,2} S(k,l) (\vec A_1 + (-1)^k \vec A_2) \otimes
(\vec B_1 + (-1)^l \vec B_2) \Big] \cdot \hat T | \le 4. \nonumber
\end{equation}

One can use the following fact. If $\vec
X_1$ and $\vec X_2$ are unit vectors, then $\vec X_1 + \vec X_2$
and $\vec X_1 - \vec X_2$ are orthogonal, and $|\vec X_1 +
\vec X_2|^2 + |\vec X_1 - \vec X_2|^2 = 4$. Therefore one can
always find two orthogonal unit vectors $\vec X(m)$, with $m=1,2$,  given by $\vec X_1 +
(-1)^m \vec X_2 = 2 x_m \vec X(m)$, such that the coefficients satisfy $\sum_m
x_m^2 = 1$. Using this convention for $\vec X_i = \vec A_i, \vec
B_i$ and $x_i=a_i,b_i$ one obtains
\begin{equation}
|\Big[ \sum_{k,l=1,2} S(k,l) a_k b_l \vec A(k) \otimes \vec B(l)
\Big] \cdot \hat T| \le 1.
\label{CONDITION}
\end{equation}
However, $[\vec A(k) \otimes \vec B(l)] \cdot \hat T = T_{kl}$ are
components of the tensor $\hat T$ in some local coordinate
system of Alice and Bob which involve vectors $\vec A(k)$ and
$\vec B(l)$ as orthogonal Cartesian coordinate directional vectors. 
Put $\vec A(1) \!= \!\hat x \!=$ $\!\hat
x_1,$ $\vec A(2) \!= $ $ \!\hat y \!=\! \hat x_2$ for
Alice's coordinate system, and similarly $\vec B(1) \!= \!\hat x
\!=$ $ \!\hat x_1,$ $\vec B(2) \!= $ $\!\hat y \!=\! \hat x_2$
for Bob's system. Thus the inequality (\ref{CONDITION}) can be expressed as $
\sum_{k,l=1,2} S(k,l) a_k b_l T_{kl} \le 1. $ 
Since one can always make the vector $(\pm a_1b_1,\pm
a_1b_2,\pm a_2 b_1, \pm a_2 b_2)$ parallel to any vector
$(T_{11},T_{12},T_{21},T_{22})$, the maximal value of the left hand side 
of the inequality (\ref{CONDITION}) 
is equal to $\max\big[{\sum_{k,l=1,2} T_{kl}^2}\big]$. Thus,
$
\max\big[{\sum_{k,l=1,2} T_{kl}^2}\big] \le 1 $ 
is the necessary and sufficient condition for the inequality 
to hold, for any measurement settings, provided the maximization
is taken over all local coordinate systems of two observers.

Consider now $4 \times 4 \times 2$
inequalities:
\begin{eqnarray}
&|\langle  \sum_{k,l} S(k,l)
(A_{12,S'}+(-1)^k A_{34,S''}) (C_1+(-1)^l C_2)
\rangle_{\textrm{avg}}| &\nonumber \\ 
&\le 16,&
\end{eqnarray}
where $S,S',S''$ are some non-factorable sign functions.
The three-qubit quantum correlation functions 
$E(\vec A_i, \vec B_j, \vec C_k)$  can be represented
as  $(\vec A_i \otimes \vec B_j
\otimes \vec C_k) \cdot \hat T$ (with the same meaning of the
symbols as before; $\hat T$ is now a three index tensor). 
Thus the
condition for the $4 \times 4 \times 2$ inequalities to hold, in
the quantum case, transforms into
\begin{equation}
|[\hat A_{12,S'} \otimes (\vec C_1 + \vec C_2) + \hat A_{34,S''}
\otimes (\vec C_1 - \vec C_2)] \cdot \hat T |\le 8, \label{KO}
\end{equation}
where e.g. \begin{equation}
\hat A_{12,S'} =\sum_{k,l=1,2} S'(k,l) (\vec A_1 + (-1)^k \vec A_2) \otimes
(\vec B_1 + (-1)^l \vec B_2). \nonumber
\end{equation}
To write down (\ref{KO}) we have used the freedom of introducing the sign changes 
$\vec X_i \to - \vec X_i$, compare (\ref{442_GEN_INEQ}).
By defining $\vec C_1 + \vec C_2 = 2 c_2 \vec C(2)$ and $\vec
C_1 - \vec C_2 = 2 c_1 \vec C(1)$ in Eq.(\ref{KO}), 
which have the properties of $\vec X_i(m)$, one obtains
\begin{equation}
|c_2 \hat A_{12,S'} \otimes \vec C(2) \cdot \hat T + c_1 \hat
A_{34,S''} \otimes \vec C(1) \cdot \hat T |\le 4.
\end{equation}
One can always choose $c_2$ and $c_1$ that maximize the left 
hand side. Since $\sum_{i=1,2} c_i^2 =1$ this leads to the condition:
\begin{equation}
[\hat A_{12,S'} \cdot \hat T^{(2)}]^{2} + [\hat A_{34,S''} \cdot
\hat T^{(1)}]^{2} \leq 4^2,
\end{equation}
where $\hat T^{(l)}$ is
defined by $T^{(l)}_{ij}=\sum_{k=1}^3 C(l)_k T_{ijk}$, where in turn
$C(l)_k$ is the $k$-th component of vector $\vec C(l)$. Note that 
since $\vec{C}(1)$ and $\vec{C}(2)$ are orthogonal and normalized
this procedure amounts to fixing of two (new) Cartesian axes for 
the third observer, and accordingly transforming the correlation tensor.
Since
$\hat A_{12,S'}$ depends on different vectors than $\hat
A_{34,S''}$, one can maximize the two terms separately. 
Furthermore, since the problem of maximization of
$\hat A_{nm,S} \cdot \hat T^{(l)}$ is equivalent to the $2 \times
2$ case studied earlier, the overall maximization process gives
the following \emph{necessary and sufficient} condition for
quantum correlations to satisfy the inequality
\begin{equation}
\max \sum_{x=1,2} \sum_{k_x,l_x=1,2} T_{k_x l_x x}^2 \le 1.
\end{equation}

When compared with the \emph{sufficient} condition for
$2 \times 2 \times 2$ inequalities to hold \cite{ZB}, namely $$
\max \big[\sum_{k,l,m=1,2} T_{klm}^2\big] \le 1,$$  the new 
condition is {\em more demanding} because the
Cartesian coordinate systems denoted by the indices $k_1,l_1$ and
$k_2,l_2$ do not have to be the same.

Consider now Bell's inequalities of the type $8 \times 8
\times 4 \times 2$. They are given by the average of
Eq.~(\ref{8842IDENTITY}) over many runs of the experiment. The
generating Bell's inequality has the form:
\begin{equation}
|\langle A_{1234,12}(D_1 + D_2) + A_{5678,34}(D_1 - D_2)
\rangle_{\textrm{avg}}| \le 32. \nonumber
\end{equation}
For the quantum predictions $E(\vec A_i,\vec
B_j, \vec C_k, \vec D_s)$,  given by 
$(\vec A_i \otimes \vec B_j \otimes \vec
C_k \otimes \vec D_s) \cdot \hat T$, the inequality  holds if
\begin{equation}
\Big( [\hat A_{1234,12} \otimes \vec D(2)] \cdot \hat T \Big)^2+
\Big( [\hat A_{5678,34} \otimes \vec D(1)] \cdot \hat T \Big)^2
\le 16^2. \nonumber
\end{equation}
The problem of maximization of either squared expression is
similar to the problem of the $4 \times 4 \times 2$ case, where now
$\hat{T}^{(l)}$ has components $T_{ijk}^{(l)}
= \sum_{r=1}^3 D(l)_r T_{ijkr}$. We can use the same maximization
algorithm as before. We get the following sufficient and
necessary condition for violation of $8 \times 8 \times 4 \times
2$ inequalities
\begin{equation}
\max \sum_{y=1,2} \sum_{x_y=1,2} \sum_{k_{xy},l_{xy}=1,2}
T_{k_{xy} l_{xy} x_y y}^2
 \le 1.
 \label{8842SN}
\end{equation}
Obviously in a similar way one can reach analogous conditions for
violation of $2^{N-1} \times 2^{N-1} \times 2^{N-2} \times ...
\times 2$ inequalities by quantum predictions.

The new conditions lead to more stringent constraints
on LR description of quantum predictions than that for the
standard inequalities. We now show some classes of states
that violate multisetting Bell's inequalities, but for which all
standard inequalities, as of refs. \cite{WZuk, WW, ZB} are satisfied. As a measure of the
violation we use the minimal amount of white noise that must
be admixed to a quantum state for a conflict with LR prediction to
disappear.

{\em Generalized GHZ-states:} $|\psi \rangle \!= \! \cos \alpha
|00...0 \rangle $ $ + \sin \alpha |11...1 \rangle $. These
states, although pure, satisfy all standard Bell's
inequalities for correlation functions for $\sin2\alpha\leq
1/\sqrt{2^{N-1}}$ and $N$ odd~\cite{SCARANI,ZBLW}. Their non
vanishing correlation tensor components are (directions $x,y,z$
are denoted by $1,2,3$; the basis $\{|0\rangle,|1\rangle\}$ is
the eigenbasis of $\sigma_z$): $T_{3...3} = \cos{2\alpha}$, for
$N$ odd, and 1 for $N$ even, $T_{1...1} = \sin{2\alpha}$, and the
components with $2k$ indices equal to $2$ and the rest equal to
$1$ take the value $(-1)^k \sin{2\alpha}$ (there are $2^{N-2}$
such components).
Let us assume that the last observer can choose only between
settings $x$ and $z$. Thus, we obtain for the condition for
violation of multisetting Bell's inequality for $N$ observers
(generalization of condition (\ref{8842SN})) \begin{eqnarray}
&\sum_{k_1,...,k_{N-1}=x,y} T^2_{k_1...k_{N-1}x}+ \sum_{k_1,
..., k_{N-1} = x,z}
T^2_{k_1...k_{N-1}z}&\nonumber \\ &= 2^{N-2} \sin^2{2\alpha} +
\cos^2{2\alpha} > 1.& \end{eqnarray}
Thus, 
the Bell's inequalities are violated for the whole range of $\pi/4\geq\alpha>0$ and for
arbitrary $N$ in contrast to the case of standard Bell's
inequalities. Note that we
did not perform any maximization.

{\em The $|W\rangle$ state:} $$1/\sqrt{N}(|100...0\rangle
+ |010...0\rangle + ... + |000...1\rangle).$$ Assume that all
$N$ observers choose between observables in the plane spanned by $y$ 
and $z$ axes. Non vanishing correlation tensor components
are $T_{z...z}=1$ and components, of modulus
equal to $2/N$, with only two indices $x$
(or $y$) and all other indices equal to $z$. Therefore the  $ \sum_{k_1,k_2,...,k_N=y,z}
T^2_{k_1...k_N} $. can be at least (no
optimization was made) $1 + {N \choose 2} \frac{4}{N^2} = 3 - 2/N > 1.$
Thus, if one considers a noise admixture to the $|W\rangle$
states, in such a form that one arrives at a mixed state
$\rho_{|W\rangle} = (1-V) \rho_{noise} + V |W \rangle \langle
W|$, with $\rho_{noise} = \openone / 2^N$, then the new
inequalities show that for $V>V_{thr}= 1/\sqrt{3- 2/N}$ there is no
LR description for correlations. This is a bigger range of $V$ than 
for the standard correlation function Bell inequalities, compare \cite{ADITI}.

Finally consider the state \cite{WZuk}:
\begin{eqnarray}
|\Psi \rangle &=& \sqrt{1/3} \Big( |0000 \rangle + |1111 \rangle \nonumber \\
&+& \frac{1}{2}(|1010\rangle + |0101 \rangle + |0110\rangle +
|1001\rangle) \Big) \\
&=&\sqrt{2/3}|\mbox{GHZ}\rangle_{1234} +
\sqrt{1/3}|\mbox{EPR}\rangle_{12}|\mbox{EPR}\rangle_{34} \nonumber
\end{eqnarray}
where $|\mbox{EPR}\rangle=1/\sqrt{2}(|01\rangle+|10\rangle)$ is
the maximally entangled (Bell-EPR) two-qubit state. This entangled
state was produced in a recent 
experiment \cite{WEINFURTERS}. Its non vanishing correlation tensor
components are: 
\begin{eqnarray}
&T_{xxxx}\!=\!T_{yyyy}\!=\!T_{zzzz}\!=\!1,& \nonumber \\
&T_{xxyy} \! = \! T_{xxzz} \! = \! T_{yyxx} \! = \! T_{yyzz}& \nonumber \\
 &= T_{zzxx} \! = \! T_{zzyy} \! = \! -1/3,& \nonumber \\
&T_{xzxz}=T_{xzzx}=T_{zxxz}=T_{zxzx}=2/3,& \nonumber \\
&T_{xyxy}=T_{xyyx}=T_{yxxy}=T_{yxyx}& \nonumber \\
&=T_{yzyz}=T_{yzzy}=T_{zyyz}=T_{zyzy}=-2/3.& \nonumber
\end{eqnarray}
The left hand side of (\ref{8842SN}) is equal to 4,  e.g. if all 
local summations are over $x$ and $y$.
Thus the $8 \times 8 \times 4 \times 2$ inequality is violated by
the factor $2$. Therefore a state $(1-V) \rho_{noise} + V |\Psi
\rangle \langle \Psi|$ gives non classical correlations for $V >
0.5$. In contrast, standard Bell inequalities cannot be
violated for $V \leq 0.5303$.

In summary, we present multipartite Bell's inequalities involving
many measurement settings and prove that they give more stringent
conditions on the possibility of a local realistic description of
quantum states, than the standard Bell's inequalities for two
settings per observer.

We thank Marcin Wie\'sniak for discussions.
The work is part of the Austrian-Polish project {\it Quantum
Communication and Quantum Information} (2004-2005) and the MNiI 
Grant No. PBZ-MIN-008/P03/2003. {\v C}.B. is supported by the Austrian FWF 
project F1506, and by the European Commission, 
Contract-No. IST-2001-38864 RAMBOQ. W.L. and T.P. are
supported by FNP. M.\.Z. is supported by the Subsydium
Profesorskie FNP. Numerical calculations were carried out at the
Academic Computer Center in Gda{\'n}sk.

\end{document}